\documentclass[]{article}
\usepackage{setspace}
\usepackage{achemso}
\usepackage{amsfonts}

\usepackage{graphicx}
\usepackage{subfigure}
\usepackage{amsmath}
\usepackage{bm}

\begin{document}
\begin{spacing}{1}

%
%
%
%
%
\def\bra#1{\mathinner{\langle{#1}|}}
\def\ket#1{\mathinner{|{#1}\rangle}}
\def\braket#1{\mathinner{\langle{#1}\rangle}}
\def\Bra#1{\left<#1\right|}
\def\Ket#1{\left|#1\right>}
{\catcode`\|=\active 
  \gdef\Braket#1{\left<\mathcode`\|"8000\let|\BraVert {#1}\right>}}
\def\BraVert{\egroup\,\mid@vertical\,\bgroup}
%


%




\title{Limits of the Plane Wave Approximation in the 
Measurement of Molecular Properties}
\author{Zachary~B.~Walters\footnote{Corresponding author: zwalters@jilau1.colorado.edu}\footnote{Department of Physics and JILA, University of
Colorado, Boulder, Colorado 80309-0440, USA} \and
Stefano Tonzani\footnote{Department of Chemistry, Northwestern University, Evanston, Illinois 60208-3113, USA} \and
Chris~H.~Greene\footnote{Department of Physics and JILA, University of
Colorado, Boulder, Colorado 80309-0440, USA}}
\date{\today}

\begin{abstract}
Rescattering electrons offer great potential as probes of molecular
properties on ultrafast timescales.  The most famous example is
molecular tomography, in which high harmonic spectra of oriented
molecules are mapped to ``tomographic images'' of the relevant
molecular orbitals.  The accuracy of such reconstructions may be
greatly affected by the distortion of scattering wavefunctions from
their asymptotic forms due to interactions with the parent ion.  We
investigate the validity of the commonly used plane wave approximation
in molecular tomography, showing how such distortions affect the
resulting orbital reconstructions.
\end{abstract}

\maketitle

When atoms or molecules are subjected to the field of an intense
laser, they may experience loss of electrons through tunnel
ionization.  The freed electrons may then propagate in the laser field
and reencounter the parent ion.  This rescattering process has been
observed to produce high harmonic generation, above threshold
ionization, nonsequential double ionization, etc., depending on which
scattering channel an experiment chooses to
monitor. \cite{corkum93,lewenstein94}  Recently,
rescattering experiments, as well as their time reversed
photoionization counterparts, have received much attention as probes of
molecular properties on ultrafast timescales
\cite{lein07,plenge06,nugentglandorf02}.  The best
known such technique is molecular tomography\cite{itatani04}, which
uses high harmonic spectra from aligned molecules to reconstruct
molecular wavefunctions.

Rescattering electrons offer clear advantages as probes of molecular
structure.  The intrinsic timescale of an
ionization-acceleration-rescattering process is on the order of a
single half-cycle of the driving laser field, typically a few fs.
Because the liberated electron is accelerated by the
driving laser field, simple formulas arising from classical physics
are sufficient to map the energy of a rescattering event to the instants
in a laser half-cycle when the electron is liberated and returns,
allowing time resolutions to be pushed to the the sub-fs level.
Interactions between electronic and vibrational degrees of freedom
permit the evolving vibrational states of the parent molecule to be
probed \cite{walters07}. 

All of these techniques rely on the same underlying physical process,
in which the rescattering electron interacts with the parent ion.
Thus, such measurements 
of molecular properties are inherently limited by the degree to which
this rescattering is understood.  However, to date most efforts to
measure molecular properties have treated the rescattering
wavefunction as a free electron plane wave, unperturbed by the electron
interaction with the 
parent ion.  Prior work relating to such reconstructions has dealt
with bandwidth limitations arising from the HHG spectrum
\cite{morishita_preprint}, orthogonality of the scattering- and bound-state
wavefunctions \cite{santra06,patchkovskii07}, and perturbative
treatments of the ionic Coulomb potential \cite{Smirnova06}.  This
paper investigates the departure from plane wave scattering which is
caused by a nonzero molecular potential and the implications of that
departure for molecular tomography.  ``Tomographic images''
of bound states are calculated for a one dimensional square well, and for
two molecules in three dimensions, $N_{2}$ and $F_{2}$.

\section{Scattering States and Ramifications for Molecular Tomography}
\label{sect:scatter}


At its heart, the tomographic procedure attempts to measure the dipole
matrix element
\begin{equation}
\label{eq:dsubk}
\vec{d}_{\vec{k}}=\int
d^{3}\vec{x}\psi_{\vec{k}}(\vec{x})\vec{x}\psi_{g}(\vec{x}) 
\end{equation}
 in momentum space between a continuum wavefunction 
$\psi_{\vec{k}}(\vec{x})$ which asymptotically
goes as $e^{i \vec{k}\cdot\vec{x}}$ and a particular orbital
$\psi_{g}(\vec{x})$ of some target molecule. In the limit where the
molecular potential is zero, the plane wave approximation for the
scattering states would be exact, and the wavefunction could be
reconstructed according to 
\begin{equation}
\label{eq:pwtomography}
\vec{x}\psi_{g}(\vec{x})=\int d^{3}\vec{k}e^{-i \vec{k}\cdot \vec{x}} 
\vec{d}_{k}(\vec{x})
\end{equation}

A nonzero molecular potential complicates this picture.  In one
dimension, the WKB approximation gives the continuum scattering state
as
\begin{equation}
\label{eq:WKB}
\psi_{c}\propto \frac{1}{\sqrt{k(x)}}e^{i\int^{x}k(x')dx'}
\end{equation}
where $k(x)=\sqrt{2(E-V(x))}$.  In the vicinity of the molecule, both
the amplitude and the phase of the scattering state depart from the
plane wave approximation.


An ideal tomographic experiment would measure $\vec{d}_{\vec{k}}$
between continuum states and an unperturbed molecular ion.  In
contrast, in rescattering experiments, recombination occurs in the
presence of a strong and time-varying external laser field.  The
magnitude of the incoming wavefunction is affected by tunnel
ionization from the molecular HOMO and the propagation of the electron
between ionization and recombination.  In addition, the high harmonic
spectrum is sharply peaked at frequencies which are multiples of the
driving laser frequency.  For these reasons, it would be very
appealing to measure $\vec{d}_{\vec{k}}$ using photoionization rather
than high harmonic generation \cite{santra06}.  It is not clear
how the phase of $\vec{d}_{\vec{k}}$ would be measured in such an
experiment, but it is at least conceivable to do so by introducing
some kind of interfering pathways.  However, since this paper is
concerned with the limitations 
to tomographic reconstruction under ideal circumstances, we henceforth
assume that such photoionization amplitudes $\vec{d}_{\vec{k}}$ could in
principle be found.  The issues discussed here with respect to tomographic
reconstruction for a photoionization experiment apply  also to HHG
tomography, with the 
stipulation that the relevant scattering states should be calculated in
the presence of an external laser field in order to provide an exact
description. 

Problems with the tomographic reconstruction procedure arise when the
scattering states $\psi_{\vec{k}}(\vec{x})$ 
begin to deviate from the plane waves
that were assumed in the initial theoretical formulations\cite{itatani04}.
As can be seen in equation \ref{eq:WKB}, this
deviation becomes pronounced when the potential experienced by the
electron is comparable 
to the scattering energy.  In this case, the measured
$\vec{d}_{\vec{k}}$ will depart from the Fourier transform of
$\vec{x}\psi_{g}(\vec{x})$.  

In equation \ref{eq:dsubk}, substitution of 
\begin{equation}
\psi_{\vec{k}}(\vec{x})=(2 \pi)^{-3/2}\int d^{3}\vec{k} 
e^{i\vec{k'}\cdot\vec{x}} \tilde{\psi}_{\vec{k}}(\vec{k'})
\end{equation}
and evaluating
the integral over $d^{3}\vec{x}$ yields 
\begin{equation}
\label{eq:distortionmap}
\vec{d}_{\vec{k}}=(2 \pi)^{-3}\int d^{3}\vec{k'}
\tilde{\psi}_{\vec{k}}(\vec{k'}) 
\widetilde{(\vec{x}\psi_{g})}_{\vec{k'}} 
\end{equation}
where $\widetilde{(\vec{x}\psi_{g})}(\vec{k'})$ represents the Fourier
transform of $\vec{x}\psi_{g}(\vec{x})$, the quantity which
tomographic procedures hope to measure, and
$\tilde{\psi}_{\vec{k}}(\vec{k'})$ represents the Fourier transform
of the scattering state $\psi_{\vec{k}}(x)$.  
In Equation \ref{eq:distortionmap}, the scattering states define a
Fourier-space mapping  from the desired function 
$\widetilde{(\vec{x}\psi_{g})}(\vec{k'})$ to the measured function
$\vec{d}_{\vec{k}}$.
In general, this mapping will not be diagonal, as the
scattering states $\psi_{\vec{k}}(\vec{x})$ will have Fourier
components at $\vec{k'}\ne\vec{k}$ due to distortions by the molecular
potential.  This mapping is not generally invertible without knowledge
of the molecular scattering states.

A 1D square well provides a simple example whose study can document
the extent to to which the electronic potential energy affects the
outcome of a tomographic reconstruction based on the plane wave
approximation.  A potential of the form
\begin{equation}
V(x)=
\left\{
\begin{array}{cc}
 V &  |x| \le x_{0} \\
 0 &   |x| > x_{0}
\end{array} \right.
\end{equation}
yields scattering states
\begin{equation}
\psi^{\text{(scat)}}_{|k|}(x)=
\left\{ \begin{array}{cc} Ae^{i k x} +Be^{-i k x} & x \le -x_{0} \\
 C e^{i k_{2} x}+De^{-i k_{2} x} & |x| \le x_{0} \\
Ee^{i k x}+Fe^{-i k x} & x > x_{0} 
\end{array} \right.
\end{equation}
where $k_{2}=\sqrt{k^{2}-2V}$.

The two linearly independent solutions $\psi_{\pm |k|}(x)$ are now chosen
such that their outgoing wave components go as $e^{\pm i k x}$ as 
$x \to \pm \infty $.  For $\psi_{|k|}$, this corresponds to 
\begin{equation}
A=\frac{4 e^{2 i (k+k_{2})x_{0}}k k_{2}}
{-(k-k_{2})^{2}+e^{4 i k_{2} x_{0}}(k+k_{2})^{2}},
\end{equation}
\begin{equation}
B=0
\end{equation}
\begin{equation}
C=\frac{2 e^{i(k+3 k_{2})x_{0}}k (k+k_{2})}
{-(k-k_{2})^{2}+e^{4 i k_{2} x_{0}}(k+k_{2})^2}
\end{equation}
\begin{equation}
D=-\frac{2 e^{i(k+k_{2})x_{0}}k(k-k_{2})}
{-(k-k_{2})^{2}+e^{4 i k_{2}x_{0}}(k+k_{2})^2}
\end{equation}
\begin{equation}
E=1
\end{equation}
\begin{equation}
F=\frac{e^{2 i k x_{0}}(e^{4i k_{2}x_{0}}-1)(k^{2}-k_{2}^{2})}
{-(k-k_{2})^{2}+e^{4 i k_{2} x_{0}}(k+k_{2})^{2}}
\end{equation}
and for $\psi_{-|k|}$,
\begin{equation}
A=\frac{e^{2 i k x_{0}}(e^{4 i k_{2}x_{0}}-1)(k^{2}-k_{2}^2)}
{-(k-k_{2})^{2}+e^{4 i k_{2}x_{0}}(k+k_{2})^{2}}
\end{equation}
\begin{equation}
B=1
\end{equation}
\begin{equation}
C=-\frac{2 e^{i(k+k_{2})x_{0}}k(k-k_{2})}
{-(k-k_{2})^{2}+e^{4 i k_{2}x_{0}}(k+k_{2})^{2}}
\end{equation}
\begin{equation}
D=\frac{2 e^{i(k+3 k_{2})x_{0}}k(k+k_{2})}
{-(k-k_{2})^{2}+e^{4 i k_{2}x_{0}}(k+k_{2})^{2}}
\end{equation}
\begin{equation}
E=0
\end{equation}
\begin{equation}
F=\frac{4e^{2i(k+k_{2})x_{0}}k k_{2}}
{-(k-k_{2})^{2}+e^{4 i k_{2}x_{0}}(k+k_{2})^{2}}
\end{equation}

Note that although
$k(x)$ takes on only two values in this problem, each scattering
solution has nonzero Fourier components for $k' \ne k, k_{2}$.

The true dipole matrix elements may now be compared to the plane wave
approximation.  A two-node bound state wavefunction was chosen as a
simple example which nonetheless possesses nontrivial spatial
structure.  Setting $x_{0}=2.5$, $V=-1.61$ yields a two-node
wavefunction with $E=-.5$.
Figure
\ref{fig:squarewelldipolecomparison}a compares dipole matrix elements
calculated 
between this bound state and continuum functions described by scattering
eigenfunctions and plane waves.  Whereas the plane wave dipole matrix
elements are all purely imaginary, calculating the matrix elements
using scattering states gives both real and imaginary components.
If the $d_{k}$ calculated using scattering states are now treated as
``measured'' dipoles and used to construct a tomographic image of the
original bound state using equation \ref{eq:pwtomography}, the
resulting image will be complex valued.  Figure
\ref{fig:squarewelldipolecomparison}b compares the original bound state
with its tomographic image.

\begin{figure}
\begin{center}
\includegraphics[width=3.375in]{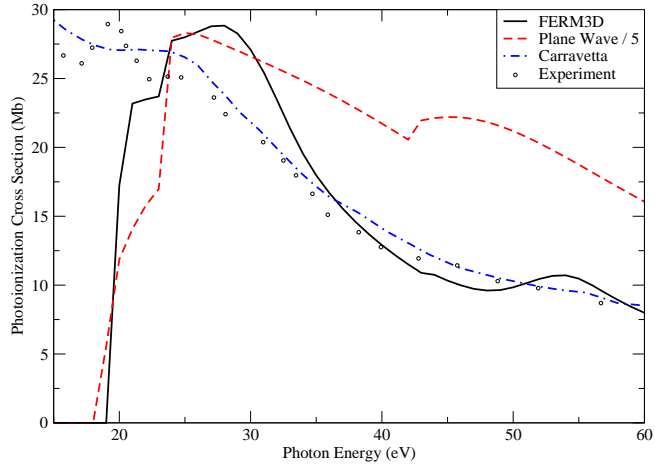}
\end{center}
\caption{(Color online) $N_{2}$ photoionization cross sections vs
  photon energy.  Calculations made using FERM3D and the plane wave
  approximation are compared to experimental measurements taken from
  \cite{carravetta93}.}
\vskip 0.45in
\label{fig:n2PIcomparison}
\end{figure}

\begin{figure}
\begin{center}
\includegraphics[width=3.375in]{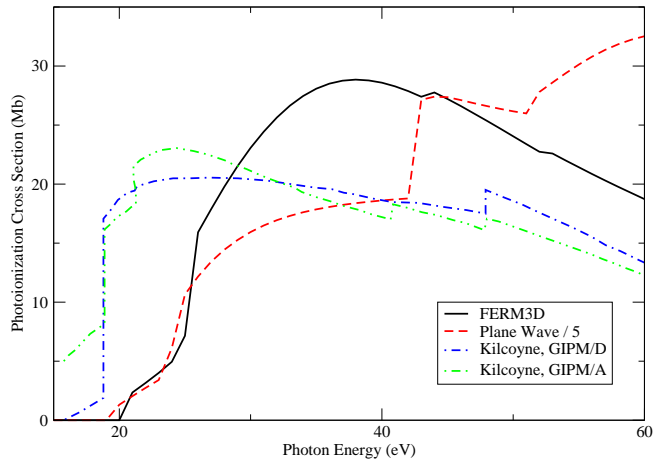}
\end{center}
\caption{(Color online) $F_{2}$ photoionization cross sections vs
  photon energy.  Calculations made using FERM3D and the plane wave
  approximation are compared to theoretical calculations taken from
  \cite{kilcoyne86}.}
\vskip 0.45in
\label{fig:f2PIcomparison}
\end{figure}

The principles seen in the case of the 1D square well also limit
tomographic reconstruction in true molecular systems, although
molecular systems are much more computationally challenging
due to the complicated potentials which describe the electron-ion
interaction.  We calculate the electron-ion scattering states using
FERM3D \cite{tonzani07}, a code which is designed for the
highly non-centrosymmetric potentials seen in these systems.  This
potential is described by 
$V_{mol}=V_{s}+V_{ex}+V_{pol}$, where $V_{s}$ is the local
electrostatic potential, $V_{ex}$ is the exchange potential arising
from antisymmetrization of the wavefunction and treated in the local
density approximation, and $V_{pol}$ is a polarization potential that
describes the relaxation of the target under the influence of the
incoming electron.  Figure \ref{fig:n2PIcomparison} compares the total
photoionization cross section for $N_{2}$ calculated with FERM3D and
the plane wave approximation to a prior calculation and
experiment \cite{carravetta93}, while Figure \ref{fig:f2PIcomparison}
compares cross sections for $F_{2}$ to a previous
calculation\cite{kilcoyne86}.  For both molecules, FERM3D gives cross
sections with sizes comparable to prior calculations, with photoionization
maxima shifted higher than in the comparison.  In both
molecules, the plane wave cross sections are too large by a factor of
five.


The calculated dipole matrix elements may also be used to
tomographically reconstruct the molecular orbitals.
In a tomography experiment, the experimentally measurable quantity is
the dipole matrix element between a bound state of the molecule and a
scattering state whose incoming-wave portion asymptotically
goes to $e^{i \vec{k}\cdot\vec{x}}$.

%
%
%
For incoming-wave boundary conditions, FERM3d calculates dipole matrix
elements
\begin{equation}
d^{q}_{E;l,m}=\bra{\psi^{(-)}_{E;l,m}}\hat{\epsilon}_{q}\cdot\vec{x}
\ket{\psi_{g}}
\end{equation}
where $\psi^{(-)}_{E;l,m}$ is an energy-normalized wavefunction which
obeys incoming-wave boundary conditions \cite{breit54}
\begin{equation}
\begin{split}
\lim_{r\to \infty}\psi^{(-)}_{E;l,m} = \sum_{l',m'}Y_{lm}(\hat{r})(2 i)^{-1}
(f_{l}^{+}(r)\delta_{l,l'}\delta_{m,m'}-\\
f_{l}^{-}(r)S^{\dag}_{l,m;l',m'}).
\end{split}
\end{equation}
where 
\begin{equation}
f^{\pm}_{l}(r)\to_{r \to \infty} e^{\pm i(kr-l\pi/2+\sigma_{l}+1/k
  \ln{2 k r})}
\end{equation}
are radially outgoing/incoming Coulomb spherical waves, as defined in
\cite{aymar96} and $\sigma_{l}=\text{arg}(\Gamma(l+1-i/k))$ is
the Coulomb phase shift.

To find $d^{q}_{\vec{k}}$, it is now necessary to find the
superposition $\psi_{\vec{k}}=A_{lm}(\vec{k})\psi_{E;lm}$ whose
outgoing component matches the outgoing component of
$e^{i\vec{k}\cdot\vec{x}}$.  Expanding
\begin{equation}
e^{i\vec{k}\cdot\vec{x}}=4\pi\sum_{l,m}i^{l}j_{l}(kr)Y_{l,m}(\hat{x})
Y^{*}_{lm}(\hat{k})
\end{equation}
where $j_{l}(kr)\to_{r \to \infty}
(2i)^{-1}(e^{i(kr-l\pi/2)}-e^{-i(kr-l\pi/2)})$ are spherical Bessel
functions, matching coefficients of $Y_{lm}(r)e^{ikr}$ yields
\begin{equation}
A_{lm}(\vec{k})=4\pi i^{l}e^{-i\sigma_{l}}Y^{*}_{lm}(\hat{k})k^{1/2}
\end{equation}
where the factor of $k^{1/2}$ converts the energy-normalized matrix
elements calculated in FERM3D to momentum normalization.

$\vec{d}_{\vec{k}}$ is now given by
\begin{equation}
\vec{d}^{q}_{\vec{k}}=\sum_{l,m}\vec{d}^{q}_{l,m}(k^{2}/2)
A_{l,m}(\vec{k}) 
\end{equation}

As in the 1D square well, the tomographic image of the orbital may
now be computed by substituting the calculated $\vec{d}_{\vec{k}}$ for
the experimentally measured quantity in the tomographic reconstruction
procedure.  As photoionization is not limited to the molecular HOMO,
tomographic images may be calculated for all the orbitals of a
molecule.  Such tomographic images will in general be complex-valued,
and will differ according to which polarization component is used for
the tomographic procedure.  

\begin{figure}
\begin{center}
\includegraphics[width=2.25in]{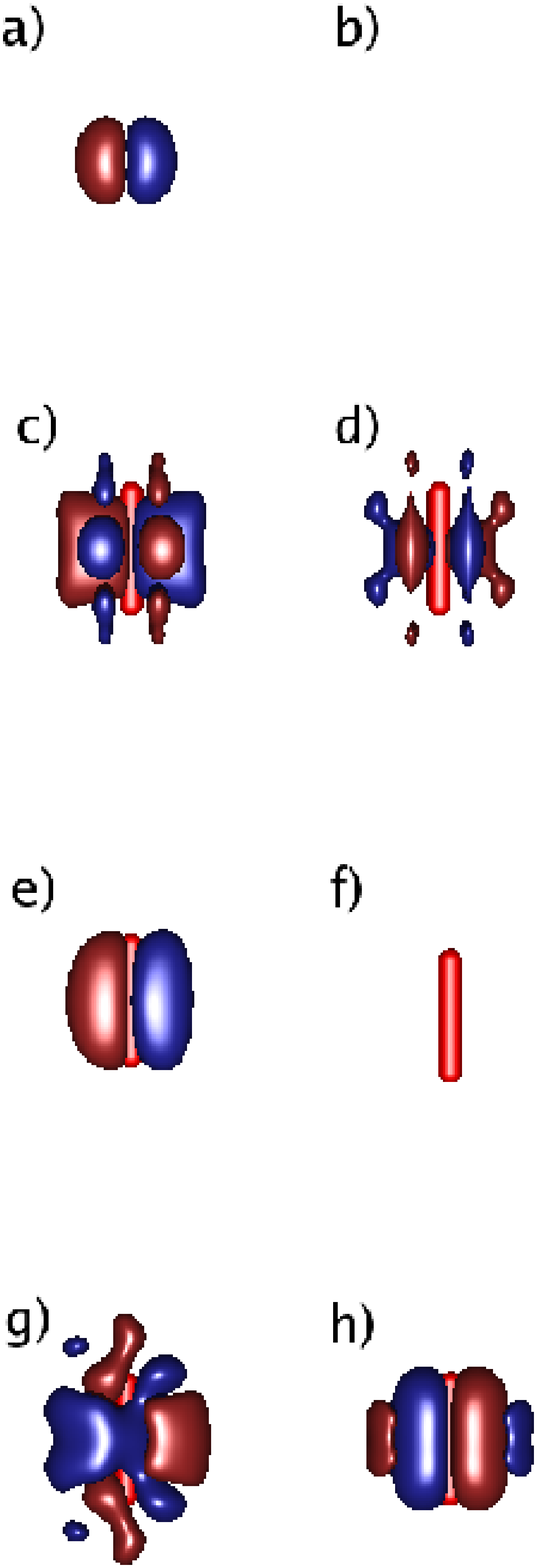}
\end{center}
\caption{(Color online) Comparison of the Hartree-Fock orbital and associated
  tomographic images for the $N_{2}$ $1\Pi_{u}$ orbital. a) and b)
  give the real and (zero) imaginary components of the Hartree-Fock
  orbital.  c) and d) give the real and imaginary components of the
  tomographic image made from the x-polarized dipole matrix element.
  e) and f) give the real and imaginary components of the tomographic
  image made from the y-polarized dipole matrix element.  g) and h)
  give the real and imaginary components of the tomographic image made
  from the z-polarized dipole matrix element.}
\label{fig:n2_1_pi_u_manyviews}
\end{figure}

\begin{figure}
\begin{center}
\includegraphics[width=2.25in]{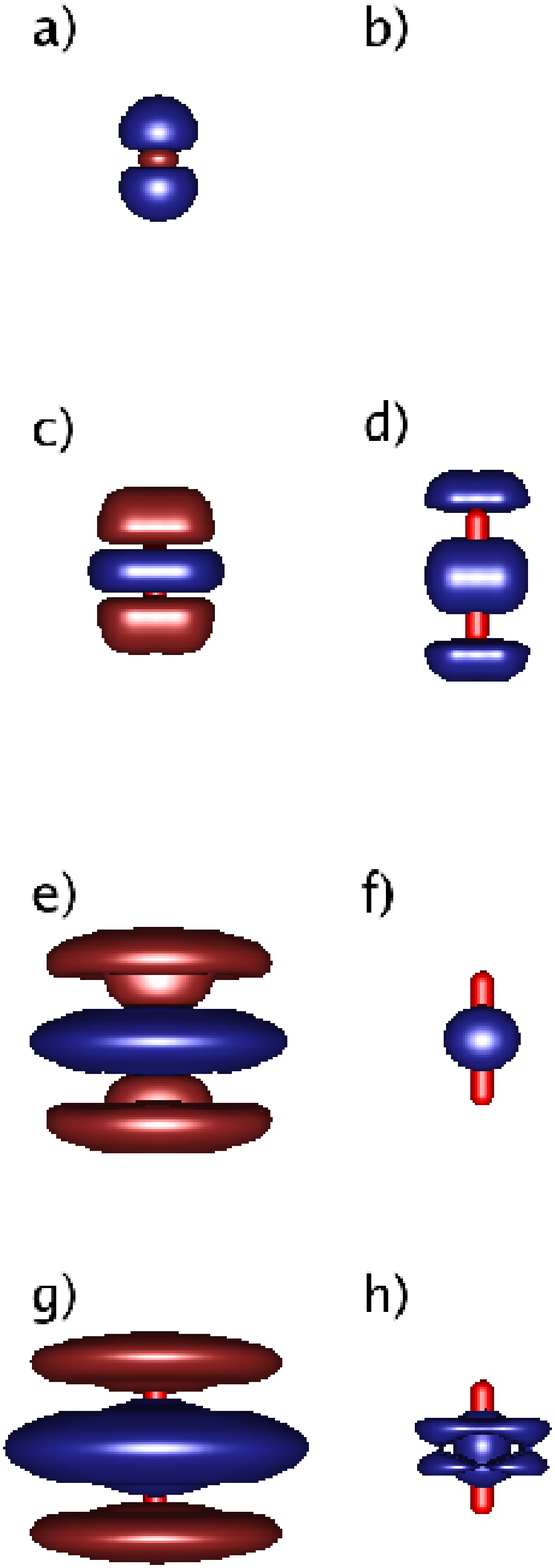}
\end{center}
\caption{(Color online) Comparison of the Hartree-Fock orbital and associated
  tomographic images for the $N_{2}$ $3\Sigma_{g}$ orbital. a) and b)
  give the real and (zero) imaginary components of the Hartree-Fock
  orbital.  c) and d) give the real and imaginary components of the
  tomographic image made from the x-polarized dipole matrix element.
  e) and f) give the real and imaginary components of the tomographic
  image made from the y-polarized dipole matrix element.  g) and h)
  give the real and imaginary components of the tomographic image made
  from the z-polarized dipole matrix element.}
\label{fig:n2_3_sigma_g_manyviews}
\end{figure}

\begin{figure}
\begin{center}
\includegraphics[width=2.25in]{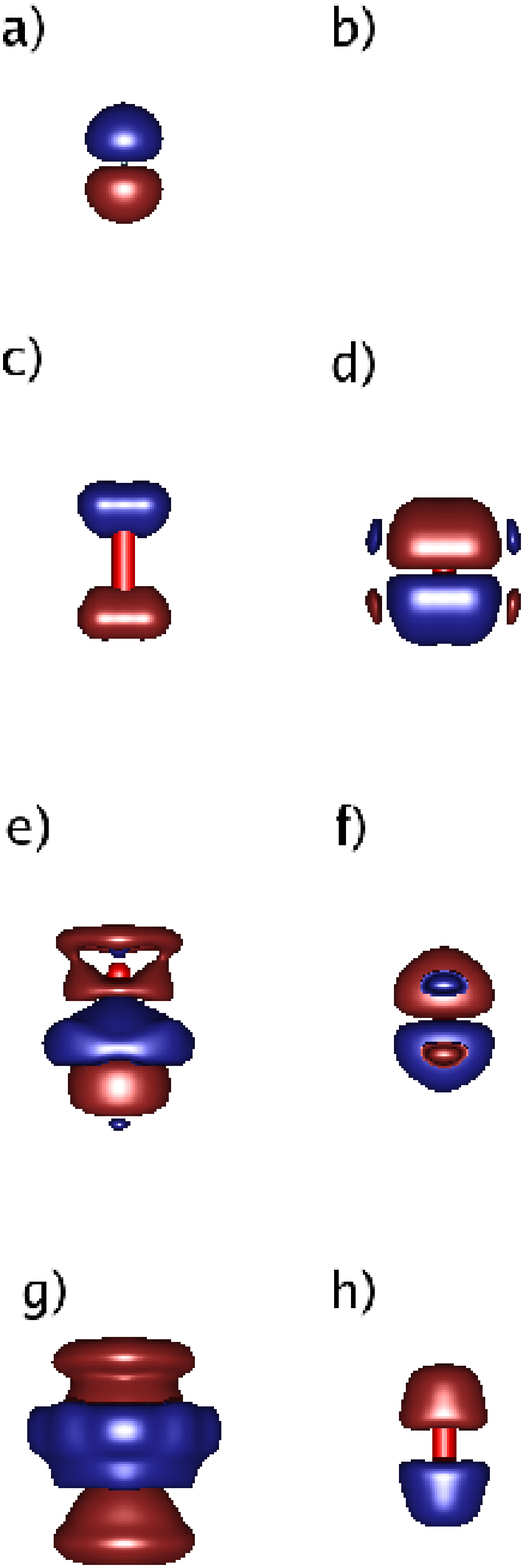}
\end{center}
\caption{(Color online) Comparison of the Hartree-Fock orbital and associated
  tomographic images for the $N_{2}$ $2\Sigma_{u}$ orbital. a) and b)
  give the real and (zero) imaginary components of the Hartree-Fock
  orbital.  c) and d) give the real and imaginary components of the
  tomographic image made from the x-polarized dipole matrix element.
  e) and f) give the real and imaginary components of the tomographic
  image made from the y-polarized dipole matrix element.  g) and h)
  give the real and imaginary components of the tomographic image made
  from the z-polarized dipole matrix element.}
\label{fig:n2_2_sigma_u_manyviews}
\end{figure}

\begin{figure}
\begin{center}
\includegraphics[width=2.25in]{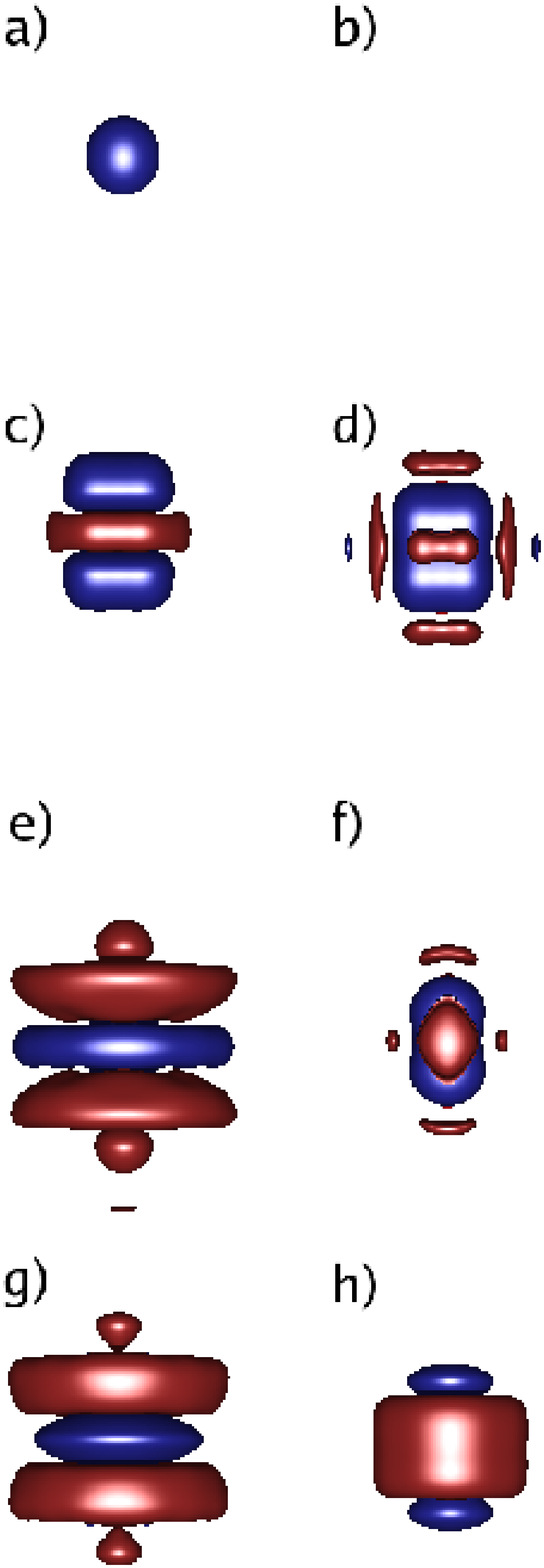}
\end{center}
\caption{(Color online) Comparison of the Hartree-Fock orbital and associated
  tomographic images for the $N_{2}$ $2\Sigma_{g}$ orbital. a) and b)
  give the real and (zero) imaginary components of the Hartree-Fock
  orbital.  c) and d) give the real and imaginary components of the
  tomographic image made from the x-polarized dipole matrix element.
  e) and f) give the real and imaginary components of the tomographic
  image made from the y-polarized dipole matrix element.  g) and h)
  give the real and imaginary components of the tomographic image made
  from the z-polarized dipole matrix element.}
\label{fig:n2_2_sigma_g_manyviews}
\end{figure}

\begin{figure}
\begin{center}
\includegraphics[width=2.25in]{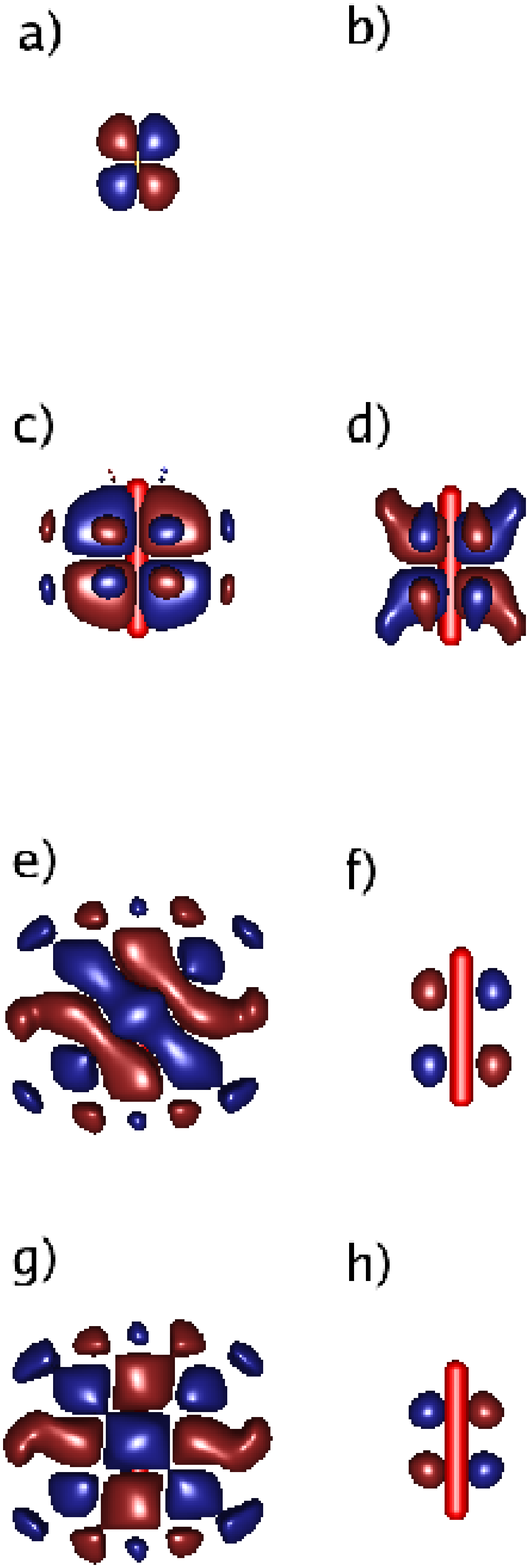}
\end{center}
\caption{(Color online) Comparison of the Hartree-Fock orbital and associated
  tomographic images for the $F_{2}$ $1\Pi_{g}$ orbital. a) and b)
  give the real and (zero) imaginary components of the Hartree-Fock
  orbital.  c) and d) give the real and imaginary components of the
  tomographic image made from the x-polarized dipole matrix element.
  e) and f) give the real and imaginary components of the tomographic
  image made from the y-polarized dipole matrix element.  g) and h)
  give the real and imaginary components of the tomographic image made
  from the z-polarized dipole matrix element.}
\label{fig:f2_1_pi_g_manyviews}
\end{figure}

\begin{figure}
\begin{center}
\includegraphics[width=2.25in]{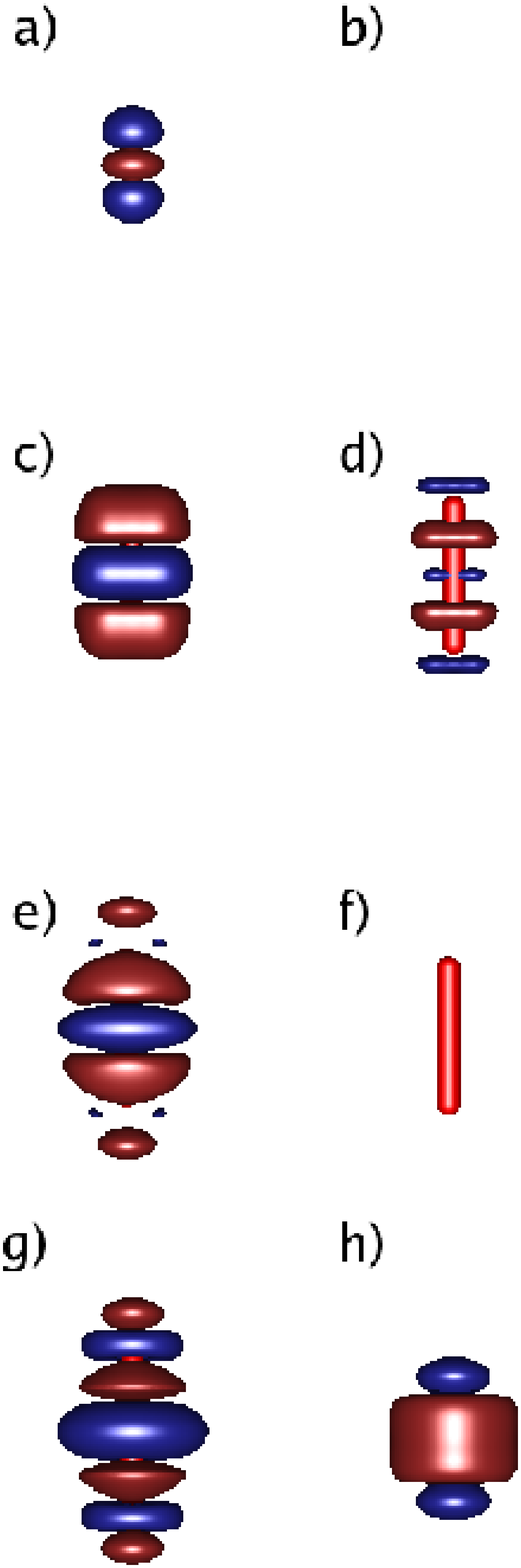}
\end{center}
\caption{(Color online) Comparison of the Hartree-Fock orbital and associated
  tomographic images for the $F_{2}$ $3\Sigma_{g}$ orbital. a) and b)
  give the real and (zero) imaginary components of the Hartree-Fock
  orbital.  c) and d) give the real and imaginary components of the
  tomographic image made from the x-polarized dipole matrix element.
  e) and f) give the real and imaginary components of the tomographic
  image made from the y-polarized dipole matrix element.  g) and h)
  give the real and imaginary components of the tomographic image made
  from the z-polarized dipole matrix element.}
\label{fig:f2_3_sigma_g_manyviews}
\end{figure}

\begin{figure}
\begin{center}
\includegraphics[width=2.25in]{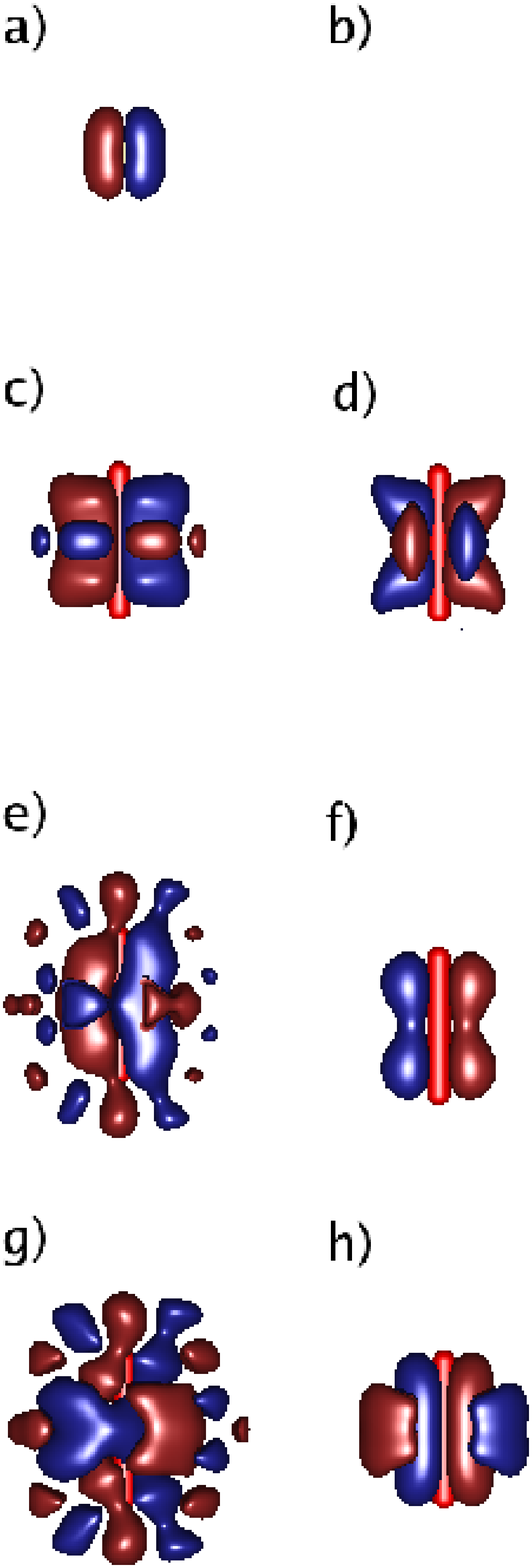}
\end{center}
\caption{(Color online) Comparison of the Hartree-Fock orbital and associated
  tomographic images for the $F_{2}$ $1\Pi_{u}$ orbital. a) and b)
  give the real and (zero) imaginary components of the Hartree-Fock
  orbital.  c) and d) give the real and imaginary components of the
  tomographic image made from the x-polarized dipole matrix element.
  e) and f) give the real and imaginary components of the tomographic
  image made from the y-polarized dipole matrix element.  g) and h)
  give the real and imaginary components of the tomographic image made
  from the z-polarized dipole matrix element.}
\label{fig:f2_1_pi_u_manyviews}
\end{figure}

\begin{figure}
\begin{center}
\includegraphics[width=2.25in]{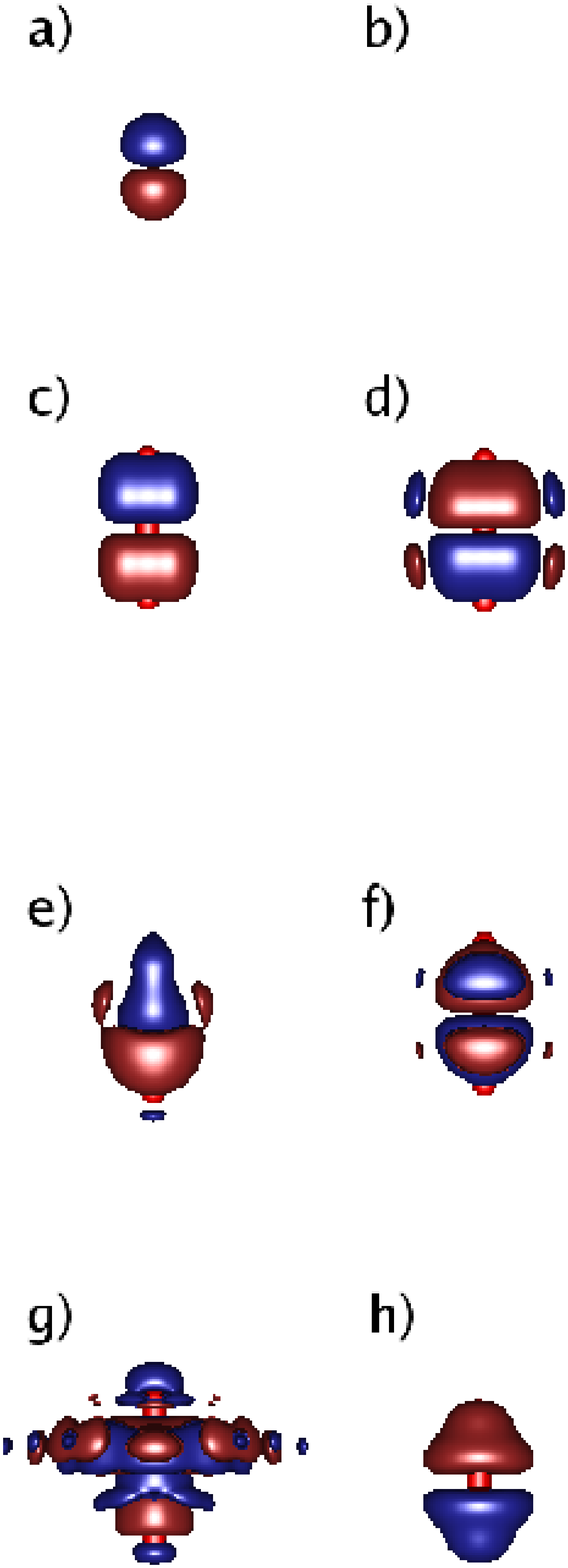}
\end{center}
\caption{(Color online) Comparison of the Hartree-Fock orbital and associated
  tomographic images for the $F_{2}$ $2 \Sigma_{u}$ orbital. a) and b)
  give the real and (zero) imaginary components of the Hartree-Fock
  orbital.  c) and d) give the real and imaginary components of the
  tomographic image made from the x-polarized dipole matrix element.
  e) and f) give the real and imaginary components of the tomographic
  image made from the y-polarized dipole matrix element.  g) and h)
  give the real and imaginary components of the tomographic image made
  from the z-polarized dipole matrix element.}
\label{fig:f2_2_sigma_u_manyviews}
\end{figure}

\begin{figure}
\begin{center}
\includegraphics[width=2.25in]{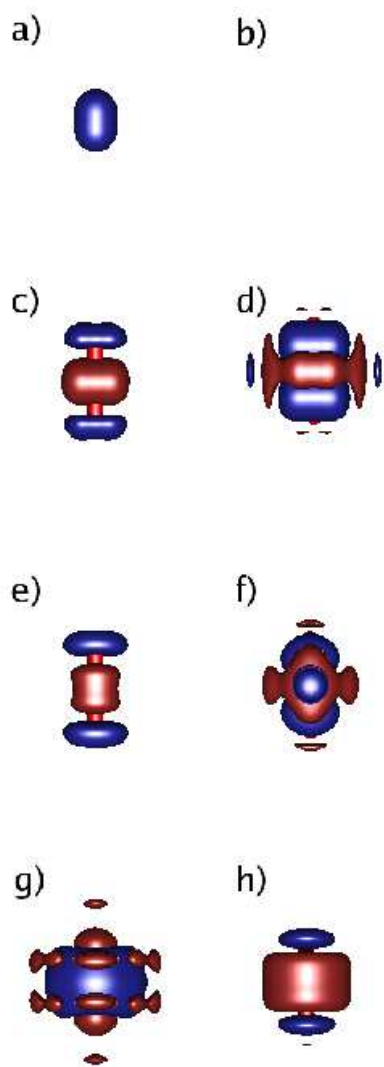}
\end{center}
\caption{(Color online) Comparison of the Hartree-Fock orbital and associated
  tomographic images for the $F_{2}$ $2\Sigma_{g}$ orbital. a) and b)
  give the real and (zero) imaginary components of the Hartree-Fock
  orbital.  c) and d) give the real and imaginary components of the
  tomographic image made from the x-polarized dipole matrix element.
  e) and f) give the real and imaginary components of the tomographic
  image made from the y-polarized dipole matrix element.  g) and h)
  give the real and imaginary components of the tomographic image made
  from the z-polarized dipole matrix element.}
\label{fig:f2_2_sigma_g_manyviews}
\end{figure}

We present 
tomographic images of various orbitals of $N_{2}$ and $F_{2}$,
calculated in the body-fixed frame using the x,y, and z polarization
components.  Each image is given with the real and imaginary
components, and includes an orange bar (not always visible) extending
between the two atoms of the molecule for the purposes of scale.
For $N_{2}$, tomographic reconstructions for the $1\Pi_{u}$
(Fig. \ref{fig:n2_1_pi_u_manyviews}), $3\Sigma_{g}$
(Fig. \ref{fig:n2_3_sigma_g_manyviews}), $2\Sigma_{u}$
(Fig. \ref{fig:n2_2_sigma_u_manyviews}) and $2\Sigma_{g}$
(Fig. \ref{fig:n2_2_sigma_g_manyviews}) orbitals are shown.
For $F_{2}$, reconstructions were calculated for the 
$1\Pi_{g}$ (Fig. \ref{fig:f2_1_pi_g_manyviews}),
$3\Sigma_{g}$ (Fig. \ref{fig:f2_3_sigma_g_manyviews}),
$1\Pi_{u}$ (Fig. \ref{fig:f2_1_pi_u_manyviews}),
$2\Sigma_{u}$ (Fig. \ref{fig:f2_2_sigma_u_manyviews}), and
$2\Sigma_{g}$ (Fig. \ref{fig:f2_2_sigma_g_manyviews}) orbitals.

For these example molecules, tomographic reconstruction tends to
preserve the $\Sigma$ or $\Pi$, gerade or ungerade character of the
orbitals in question.  However, the reconstructed orbitals may display
additional radial nodes not found in the original orbitals.  Features
which correspond to features of the original orbitals may be distorted
in shape and size, and display a spatially varying complex phase.
Finally, tomographic images of the same orbital made using different
polarization information may produce differing images of the same
orbital.  Many of these features are also seen in
\cite{patchkovskii07}, which treats the scattering process from a
multielectron perspective, but does not consider the distorting
effects of a molecular potential.

\section{Conclusions}
%

The use of rescattering electrons as a probe of molecular properties
offers many exciting avenues for future research.  However, the
rescattering process is itself more complicated than has been
recognized in early reconstruction efforts, and is worthy of study in
its own right.

For molecular tomography, the results presented in this paper suggest
that at energy scales where such distortion is significant,
tomographic reconstructions may be significantly distorted from the
``true'' orbitals these methods seek to find.  Nevertheless, for the
example molecules presented here, the tomographic reconstruction
procedure was able to successfully reproduce the $\Sigma$ or $\Pi$, gerade or
ungerade nature of the orbitals in question.  Reconstructions made
using differently polarized dipole matrix elements gave different
tomographic images of the same orbital, while reconstructions made
from a particular polarization gave tomographic images with spatially
varying complex phase.  Both of these properties could be useful as an
experimental check of reconstructed wavefunctions: ideal
reconstructions would have a spatially uniform complex phase and
reconstructions made with different polarization information should
agree with one another.  Additionally, experiments could use higher
scattering energies to minimize the scattering state distortions due
to interactions with the molecular potential.

The sensitivity of the scattering states to the molecular potential
also offers the prospect for new types of experiments.  Such
experiments could monitor the movement of charge within the parent ion
at ultrafast timescales.  For example, a two-center interference
experiment of the type discussed in \cite{lein07} might observe the
movement of charge in a diatomic molecule by observing interference
maxima/minima occurring at different energies for the short and long
rescattering trajectories.

We thank Nick Wagner for insightful and stimulating discussions.
This work was supported in part by the Department of Energy, Office of
Science, and in part by the NSF EUV ERC.

\section{Appendix: Gauges and Dispersion Relations}
Within the overall framework of the plane wave approximation, several
heuristic methods have been suggested to improve the accuracy of
tomographic reconstructions\cite{lein07}.  For the 1D square well, we
considered the effects of phenomenological ``dispersion relations''
and reconstructions made in the momentum, rather than velocity gauge.

A dispersion relation attempts to correct for an electron's shorter
wavelength by substituting $e^{i \vec{q}\cdot\vec{x}}$ for 
$e^{i \vec{k}\cdot\vec{x}}$ in Equation \ref{eq:pwtomography}, where 
$|\vec{q}|=\sqrt{2(k^{2}/2- \epsilon V)}$, where $V$ is the potential
felt by the electron in the interaction region, and $\epsilon \in
[0,1]$.

Tomography may also be performed in gauges other than the length gauge
given in equation \ref{eq:pwtomography}.  If both continuum and bound
wavefunctions are eigenstates of the Hamiltonian, the dipole matrix
element is identical in the length
\begin{equation}
d^{(l)}_{k}=\bra{\psi_{g}}x\ket{\psi_{k}}
\end{equation}
and momentum
\begin{equation}
d^{(p)}_{k}=\frac{\bra{\psi_{g}}ip\ket{\psi_{k}}}{E_{g}-E_{k}}
\end{equation}
gauges.  From the momentum gauge form of the dipole matrix element,
and employing the plane wave approximation for $\ket{\psi_{k}}$, it
is possible to generate a second tomographic reconstruction
\begin{equation}
\psi^{(p)}(x)=\int dq e^{iq(k)x}d^{(p)}_{k} \frac{E_{g}-k^{2}/2}{-q(k)}.
\end{equation}
As a plane wave is not an eigenfunction of the scattering Hamiltonian,
this reconstruction will in general give a different image of the
target orbital than a reconstruction made using the length gauge.  

\begin{figure}
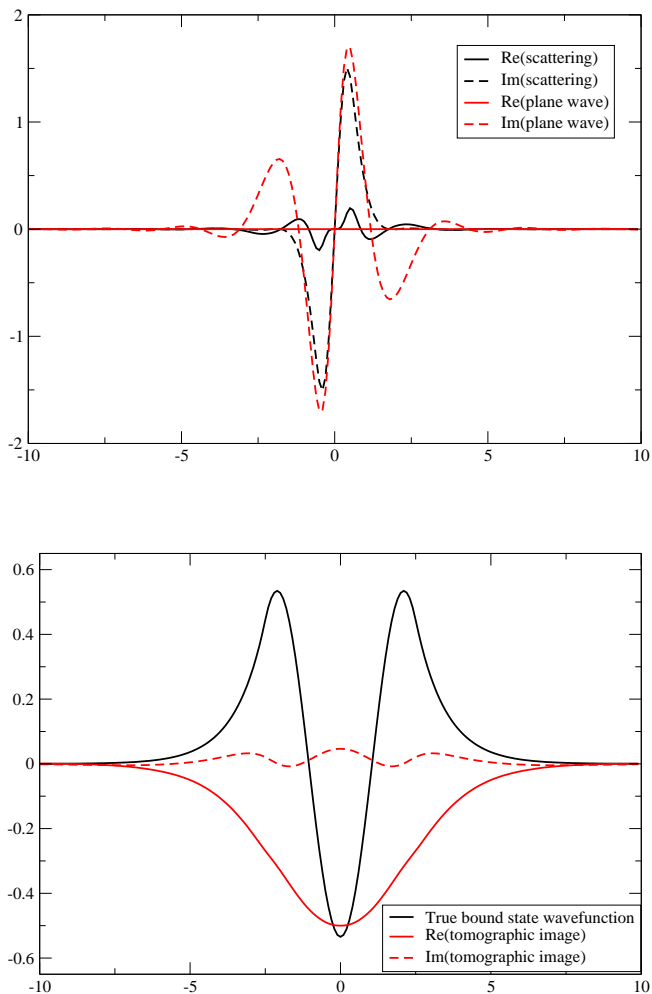

\begin{center}
\includegraphics[width=3.375in]{Squarewelltomography_dipole_comparison_nobarrier.eps}
\vskip 0.45in
\includegraphics[width=3.375in]{Squarewell_tomographic_image_nobarrier.eps}
\end{center}
\caption{(Color online) The tomographic reconstruction procedure
  applied to the 1D square well.  (Top) Comparison of dipole matrix
  elements $d_{k}=\bra{\psi_{c}(x)}x\ket{\psi_{g}(x)}$, calculated using
  plane waves and scattering states for $\psi_{c}(x)$.
 (Bottom) Because the scattering state matrix elements differ from
  those calculated using plane waves, the reconstructed image of the
  orbital will differ from the true bound state wavefunction.}
\vskip 0.45in
\label{fig:squarewelldipolecomparison}
\end{figure}

We tested both the length- and momentum-gauge tomographic
reconstructions using dispersion relations 
$q(\epsilon)=\sqrt{2(k^{2}/2-\epsilon V)}$, using
the overlap of the true wavefunction and its (normalized) tomographic
image as a figure of merit.  The test employed the same $V=-1.61$, $x_{0}=2.5$
potential and $E=-.5$ target wavefunction used to generate Figure
\ref{fig:squarewelldipolecomparison}.

\begin{figure}
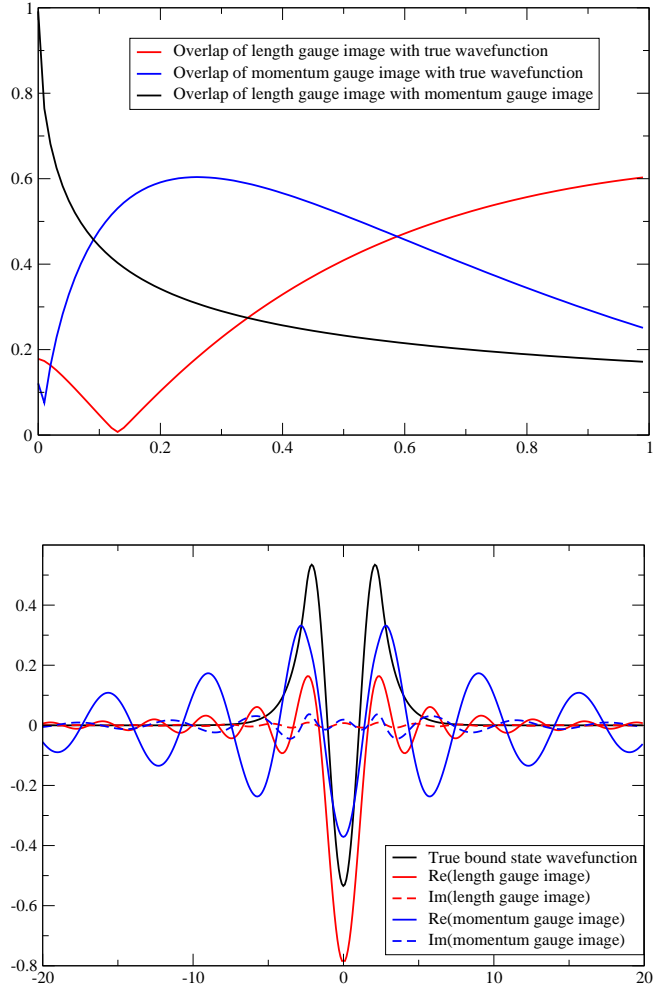

\begin{center}
\includegraphics[width=3.375in]{tomographic_image_overlap_vs_dispersion_relation.eps}
\vskip 0.45in
\includegraphics[width=3.375in]{maxolap_tomographic_image_nobarrier.eps}
\end{center}
\caption{(Color online) a) Magnitude of the overlap between (normalized)
  tomographic images of a bound wavefunction and the true wavefunction, and
  between different tomographic images, calculated using
  $q(\epsilon)=\sqrt{2(k^{2}/2-\epsilon V)}$, $\epsilon \in [0,1]$.
  b) Comparison of the maximally overlapping tomographic images to the
  true wavefunction.  In the momentum gauge, maximal overlap was
  obtained for $\epsilon=.26$, while in the length gauge, maximal
  overlap was obtained for $\epsilon=1$.  Both images have been
  normalized and rotated to give a purely real overlap with the true
  wavefunction.}
\vskip 0.45in
\label{fig:gaugedispersionoverlap}
\end{figure}

Figure \ref{fig:gaugedispersionoverlap}a gives the magnitude of the
overlap between the two tomographic images and the ground state and
between each other as a function of the dispersion parameter $\epsilon$.  
For this choice of potential and target orbital, the tomographic
reconstructions gave a very poor overlap with the target orbital for
$\epsilon \in [0,1]$, reaching a maximum magnitude of $0.60$ at
$\epsilon=0.26$ in the momentum gauge.  In the dipole gauge, the
maximum overlap was achieved at $\epsilon=1.0$, also giving an overlap
of magnitude 0.60.  Figure \ref{fig:gaugedispersionoverlap}b compares
the maximally overlapping reconstructions to the true ground state
wavefunction.

A perfect reconstruction  would give the same image regardless of the
gauge the tomographic procedure was performed in.  However, agreement
between images made in separate gauges does not guarantee the accuracy
of the reconstruction.  Although both gauges gave nearly identical
tomographic images at $\epsilon=0$, the resulting images gave among
the worst overlaps with the target orbital.  



\begin{thebibliography}{10}

\bibitem{corkum93}
Corkum,~P.~B. \textit{Phys. Rev. Lett.} \textbf{1993,} \textsl{71,} 1994.

\bibitem{lewenstein94}
Lewenstein,~M.;\ \ Balcou,~P.;\ \ Ivanov,~M.~Y.;\ \ A.L'Huillier,;\ \
  Corkum,~P.~B. \textit{Phys. Rev. A} \textbf{1994,} \textsl{49,} 2117.

\bibitem{lein07}
Lein,~M. \textit{J. Phys. B} \textbf{2007,} \textsl{40,} R135.

\bibitem{plenge06}
Plenge,~J.;\ \ Nicolas,~C.;\ \ Caster,~A.~G.;\ \ Ahmed,~M.;\ \ Leone,~S.~R.
  \textit{J. Chem. Phys.} \textbf{2006,} \textsl{125,} 133315.

\bibitem{nugentglandorf02}
Nugent-Glandorf,~L.;\ \ Scheer,~M.;\ \ Samuels,~D.~A.;\ \ Bierbaum,~V.~M.;\ \
  Leone,~S.~R. \textit{J. Chem. Phys.} \textbf{2002,} \textsl{117,} 6108.

\bibitem{itatani04}
Itatani,~J.;\ \ Levesque,~J.;\ \ Zeidler,~D.;\ \ Niikura,~H.;\ \ Pepin,~H.;\ \
  Kieffer,~J.~C.;\ \ Corkum,~P.~B.;\ \ Villeneuve,~D.~M. \textit{Nature}
  \textbf{2004,} \textsl{432,} 867.

\bibitem{walters07}
Walters,~Z.;\ \ Tonzani,~S.;\ \ Greene,~C.~H. \textit{J. Phys. B.}
  \textbf{2007,} \textsl{40,} F277.

\bibitem{morishita_preprint}
Morishita,~T.;\ \ Le,~A.~T.;\ \ Chen,~Z.;\ \ Lin,~C.~D. \textit{preprint}  .

\bibitem{santra06}
Santra,~R.;\ \ Gordon,~A. \textit{Phys. Rev. Lett.} \textbf{2006,} \textsl{96,}
  073906.

\bibitem{patchkovskii07}
Patchkovskii,~S.;\ \ Zhao,~Z.;\ \ Brabec,~T.;\ \ Villeneuve,~D.~M. \textit{J.
  Chem. Phys.} \textbf{2007,} \textsl{126,} 114306.

\bibitem{Smirnova06}
Smirnova,~O.;\ \ Spanner,~M.;\ \ Ivanov,~M. \textit{J. Phys. B} \textbf{2006,}
  \textsl{39,} S307.

\bibitem{tonzani07}
Tonzani,~S. \textit{Comp. Phys. Comm.} \textbf{2007,} \textsl{176,} 146.

\bibitem{carravetta93}
Carravetta,~V.;\ \ Luo,~Y.;\ \ Agren,~H. \textit{Chem. Phys.} \textbf{1993,}
  \textsl{174,} 141.

\bibitem{kilcoyne86}
Kilcoyne,~D.~A.~L.;\ \ Nordholm,~S.;\ \ Hush,~N.~S. \textit{Chemical Physics}
  \textbf{1986,} \textsl{107,} 197.

\bibitem{breit54}
Breit,~G.;\ \ Bethe,~H.~A. \textit{Phys. Rev.} \textbf{1954,} \textsl{93,} 888.

\bibitem{aymar96}
Aymar,~M.;\ \ Greene,~C.~H.;\ \ Luc-Koenig,~E. \textit{Rev. Mod. Phys.}
  \textbf{1996,} \textsl{68,} 1015.

\end{thebibliography}
\end{spacing}

\end{document}